\documentclass[aps,prb,twocolumn,showpacs,superscriptaddress]{revtex4}
\usepackage[latin1]{inputenc}
\usepackage[english]{babel}
\usepackage{bm,graphicx,graphics,amsmath,amssymb,epsfig,color}

\usepackage[colorlinks,citecolor=blue,linkcolor=cyan]{hyperref}
\usepackage[usenames,dvipsnames,svgnames,table]{xcolor}

\def \be {\begin{equation}}
\def \ee {\end{equation}}
\def \beA {\begin{eqnarray}}
\def \eeA {\end{eqnarray}}

\def \average#1{\left\langle #1 \right\rangle}

\def \graffb#1{\left\{ #1 \right\}}

\begin{document}

\title{Nanoscale control of heat and spin conduction in artificial spin chains}

\author{Simone Borlenghi} 
\affiliation{Department of Physics and Astronomy, Uppsala University, Box 516, SE-75120 Uppsala, Sweden.}
\author{M. R. Mahani}
\affiliation{Department of Materials and Nanophysics,  School of Information and Communication Technology, \\Electrum 229, Royal Institute of Technology, SE-16440 Kista, Sweden.}
\author{Anna Delin}
\affiliation{Department of Physics and Astronomy, Uppsala University, Box 516, SE-75120 Uppsala, Sweden.}
\affiliation{Department of Materials and Nanophysics,  School of Information and Communication Technology, \\Electrum 229, Royal Institute of Technology, SE-16440 Kista, Sweden.}
\affiliation{Swedish e-Science Research Center (SeRC), KTH Royal Institute of Technology, SE-10044 Stockholm, Sweden}
\author{Jonas Fransson}
\affiliation{Department of Physics and Astronomy, Uppsala University, Box 516, SE-75120 Uppsala, Sweden.}

\begin{abstract}
We describe a mechanism to control energy and magnetisation currents in an artificial spin-chain, consisting of an array of Permalloy nano-disks coupled through the magneto-dipolar interaction.
The chain is kept out of equilibrium by two thermal baths with different temperatures connected to its ends, which control the current propagation.
Transport is enhanced by applying a uniform radio frequency pump field resonating with some of the spin-wave modes of the chain. 
Moreover, the two currents can be controlled independently by tuning the static field applied on the chain. 
Thus we describe two effective means for the independent control of coupled currents and the enhancement of thermal and spin-wave conductivity in a realistic magnonics device, suggesting that similar effects could be observed in a large class of nonlinear oscillating systems.
\end{abstract}

\maketitle 

\newpage

\section{Introduction} %
Finding ways for the nanoscale control of coupled currents in systems with several conserved quantities is of primary importance for applications in nanoscale devices for energy
harvesting, nano-phononics~\cite{lepri03,dhar08,balandin12}, thermoelectric and thermomagnetic conversion~\cite{casati08}.
These fields have attracted a lot of attention, undergoing an unprecedented development due to recent discoveries in spin-caloritronics~\cite{uchida08,uchida10,bauer12}. 
In magnetic nano-structures, thermal gradients can be used to propagate energy and magnetisation (or spin-wave, SW) currents, controlling several non-equilibrium transport phenomena \cite{xiao10,yu10,hinzke11}

Spintronics and magnonics devices \cite{kruglyak10} are promising for applications especially owing to versatility. In fact, they can be controlled by a combination of various means, such as thermal gradients, magnetic fields and electrical currents~\cite{wolf01,zutic04}. 
On a more fundamental level, they constitute simple setups that allow to study quite deep and general problems of non-equilibrium thermodynamics, notably the connection between phase-coherence and transport \cite{borlenghi14a}.

Theoretically, this issue can be addressed using the language of the non-equilibrium discrete nonlinear Schr\"odinger (DNLS) equation~\cite{iubini12,borlenghi14a,borlenghi15a,borlenghi15b,borlenghi16a}.
The latter is a general oscillator model with applications in several branches of  physics, including spin systems \cite{slavin09}, Bose-Einstein condensates, photosyntetic reactions \cite{iubini15}, lasers and mechanical oscillators~\cite{Kevrekidis}.
Our predictions should be possible to test using networks of nanoscale ferromagnetic islands. Such networks have been successfully used in recent years to experimentally study, e.g., magnetic charges and magnetic frustration \cite{wang06,morgan11,zhang13}.

\begin{figure}[t]
\begin{center}
\includegraphics[width=9cm]{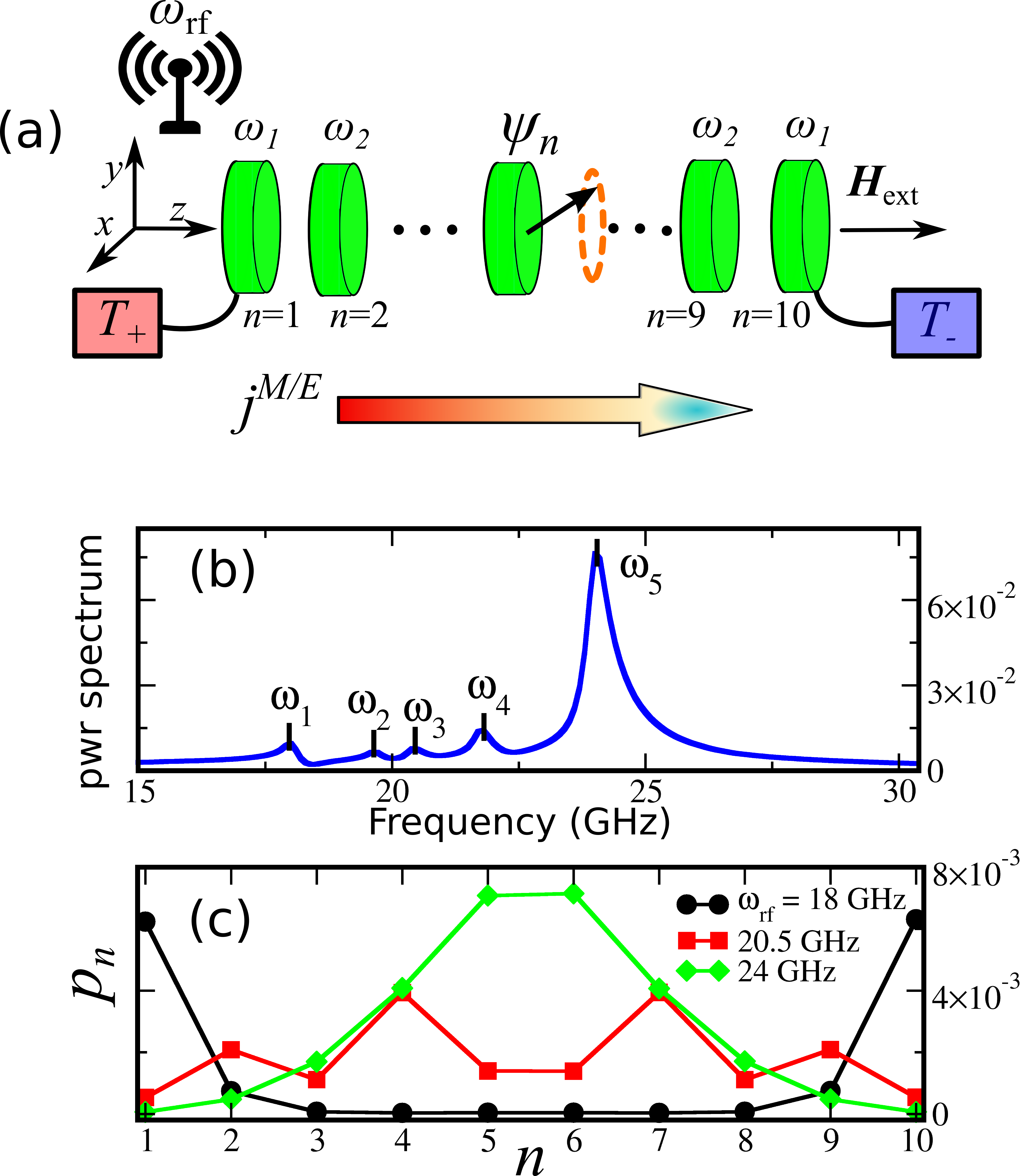}
\end{center}
\caption{a) Chain of disks coupled via the magneto-dipolar interaction. Each disk behaves as a precessing macrospin with frequency $\omega_n$. The first and last disks are 
coupled to thermal baths with temperatures $T_\pm$ respectively, which control the propagation of the magnetisation and energy currents $j_n^{M/E}$. 
Transport is enhanced by applying a uniform rf field with intensity $h_{\rm{rf}}$ and frequency $\omega_{\rm{rf}}$, which increases both the coherence and the local precession amplitudes. 
The applied static field $H_{\rm{ext}}$ makes it possible to control the two currents separately. b) SW spectrum of the system, consisting of five localised dipolar modes (see text). c) SW power profiles at zero temperature, obtained
exciting the dynamics with uniform rf fields with frequencies $\omega_{\rm{rf}}=\omega_1,\omega_3$ and $\omega_5$. The lines are guides to the eye.}
\label{fig:figure1}
\end{figure}

In this Paper, we describe a mechanism to control independently energy and magnetisation currents in an artificial spin-chain. 
Our setup, shown in Fig.\ref{fig:figure1}a),  consists of an array of ten Permalloy (Py) nano-disks coupled through the magneto-dipolar interaction, with the first and last disks attached to stochastic baths with temperatures $T_{\pm}$, respectively. In the presence of a temperature difference $\Delta T=T_+-T_-$, the chain reaches a non-equilibrium steady state where energy and magnetisation currents flow from the hot  ($T_+$) to the cold ($T_-$) reservoirs.

In the presence of a thermal gradient, transport can be controlled by applying a uniform radio frequency (rf) field resonating with some of the SW modes of the system. The rf field acts by increasing the phase-locking between the oscillators,
and by exciting the dynamics at the edges of the chain. In this way, the spin-chemical potential at the boundaries of the system grows and both currents increase.
Moreover, currents can be controlled independently by tuning the static magnetic field applied on the chain. At increasing field, we find that the magnetisation current is suppressed while the energy current remains unchanged.
These properties can be qualitatively understood using the DNLS model.

The present Paper is organised as follows: In Sec. II we describe the system and we briefly review the transport properties of the DNLS chain. In Sec. III we investigate the dynamics of the system by means of micromagnetic simulations, focussing in 
particular on the interplay of thermal gradient and rf pump field to control heat and spin transport. Finally, in Sec. IV we resume the main results of the paper.

\section{Physical system and model}%

Let us start by a brief review of the dynamics of the chain and the DNLS model. We refer to Ref.~[\onlinecite{borlenghi15b}] for a thorough discussion.
The coherent dynamics of the magnetisation in the $n_{\rm{th}}$ disk is described in terms of a single magnetic moment $\bm{M}^n=M_s{\bm{m}}^n$, with constant length $M_s$ and direction $\bm{m}^n$,
which obeys a  Landau-Lifschitz-Gilbert (LLG) type equation~\cite{borlenghi15b}:

\be\label{eq:llg}
\dot{\bm{M}}_{n}=\gamma(\bm{H}_{\rm{eff}}\times\bm{M}_n)+\alpha(\bm{M}_n\times\dot{\bm{M}}_n).
\ee
The first term of Eq.(\ref{eq:llg}) describes the precession of the magnetisation around the effective field $\bm{H}_{\rm{eff}}$, while the second term accounts for energy dissipation at a rate proportional to the Gilbert damping parameter
$\alpha$.

The effective field consists of the sum of the following terms $\bm{H}_{\rm{eff}}=\bm{H}_{\rm{ext}}+\bm{H}_{\rm{dip}}+\bm{H}_{\rm{rf}}+\bm{H}_{\rm{th}}$. Those are respectively the external field along $\bm{z}$, which defines the precession axis of the 
magnetisation, the dipolar field responsible for the interlayer coupling, the applied rf field used to excite the dynamics, and the thermal field. The latter describes thermal fluctuations in the $n$th  disk in terms of a white noise process with statistical properties

\beA\label{eq:hth}
\average{\bm{H}_{n\rm{th}}}(t) &=& 0\nonumber,\\
\average{\bm{H}_{n\rm{th}}(t)\bm{H}_{n^\prime\rm{th}}(t^\prime)} &=& \frac{2\alpha K_{B}T_{n}}{\gamma V M_s}\delta_{nn^\prime}\delta(t-t^\prime),
\eeA
where $k_B$ is the Boltzmann constant,  while $V_n$ and $T_n$ are respectively the volume and temperature of the $n$th disk.

The small amplitude dynamics of the chain (with polar angle $\theta_n\leq10^\circ$) is conveniently expressed in terms of the complex spin wave (SW) amplitude~\cite{slavin09,borlenghi15b}

\be\label{eq:psi}
\psi_n=\frac{M_x^n+iM_y^n}{\sqrt{2M_s(M_s+M_z^n)}}\equiv\sqrt{p_n(t)}e^{i\phi_n(t)}, 
\ee
where both the SW powers $p_n(t)$ and $\phi_n(t)$ are time dependent.
The SW power $p_n=|\psi_n|^2$ is related to the polar angle of the magnetisation $\theta_n=\arccos(1-2p_n)$, while $\phi_n(t)$ describes the precession of $\bm{M}^n$ in the $x$-$y$ plane.

In terms of the $\psi_n$s, the LLG equation (\ref{eq:llg}) transforms into the DNLS \cite{slavin09,iubini13,borlenghi15b}:
\beA\label{eq:dnls}
i\dot{\psi}_n & = & -\omega_n(p_n)\psi_n-i\Gamma_n(p_n)\psi_n\nonumber\\
		   & - &J(1+i\alpha)(\psi_{n+1}+\psi_{n-1}) +h_{\rm{rf}}e^{i\omega_{\rm{rf}}t}\nonumber\\
                   & + &  \sqrt{D_n(p_n)T_n}\xi_n.
\eeA
The first two terms on the right hand side of Eq.(\ref{eq:dnls}) are respectively the nonlinear frequencies $\omega_n(p_n)=\omega_n^0(1+Ap_n)$ and damping rates $\Gamma_n(p_n)=\alpha\omega_n^0(1+Bp_n)$ of the $n_{\rm{th}}$ disk.
Here $A$ and $B$ are the coefficients of the expansion of $\omega_n(p_n)$ and $\Gamma_n(p_n)$  to the first order in $p_n$.
The frequencies $\omega_n^0$ are proportional to $\gamma H_{\rm{ext}}$, where $\gamma$ is the gyromagnetic ratio. 
The third term $J$ is the strength of the magneto-dipolar coupling. Although this coupling decreases as a function of distance with a power law, a direct comparison with the numerical integration of the DNLS has shown that it can be approximated by a nearest neighbour interaction \cite{borlenghi15b}.  We note that, because of the fluctuation-dissipation theorem, the coupling constant $J$ has to be multiplied by the factor $(1+i\alpha)$ \cite{iubini13,borlenghi15b}. This condition, called dissipative coupling,
ensures that the system reaches thermal equilibrium when baths have the same temperature.
The dipolar field is also responsible for the nonlinearity of the DNLS Eq.(\ref{eq:dnls})~\cite{slavin09} and for the presence of localised SW modes with five different frequencies. 
The quantity $\sqrt{D_n(p_n)T_n}\xi_n$ models thermal fluctuations in terms of the complex Gaussian random variables $\xi_n$, with statistical properties
\beA
\average{\xi_n(t)} &=& 0\nonumber\\
\average{\xi_n(t)\xi^*_{n^\prime}(t^\prime)} &=& \delta_{nn^\prime}\delta(t-t^\prime).  
\eeA
The fluctuation-dissipation theorem fixes the coupling strength with the bath: 
\be\label{eq:fdt}
D_n(p_n)=\frac{\Gamma_n(p_n)}{\lambda\omega_n(p_n)}, 
\ee
with $\lambda$ a coefficient that depends on the geometry of system~\cite{slavin09}. 
The last term of Eq.(\ref{eq:dnls}) describes the effect of a uniform rf pump field polarised in the $x$-$y$ plane, with intensity $h_{\rm{rf}}$ and frequency $\omega_{\rm{rf}}$. The parameters $(A,B,J)$ can be estimated analytically only in some simple cases and are usually inferred from micromagnetic simulations~\cite{slavin09,naletov11}.

Note that, in the small amplitude regime, the LLG Eq.(\ref{eq:llg}) and the DNLS Eq.(\ref{eq:dnls}) are completely equivalent. The advantage of adopting the DNLS language here is twofold. On one hand, it allows for a simple expression of the dipolar field, which is otherwise rather complicated \cite{borlenghi15b}. On the other hand, it make the description of coupled transport more transparent and simplifies the expression of energy and SW currents.
Moreover, mapping the LLG equation into the more general DNLS formalism serves to elucidate features of magnetic systems that could be found in other branches of Physics, in particular the connection between transport and synchronisation. 

This mapping works when the small nonlinearity in the DNLS can be expressed in powers of $p_n$. In principle one can use the transformation Eq.(\ref{eq:psi}) for an arbitrary large precession angle of the magnetisation. However In this case 
one obtains a more complicated equation than the DNLS [REFS].


The conservative part of the DNLS Eq.(\ref{eq:dnls}) is obtained from the Hamilton equation $\dot\psi_n=-\partial\mathcal H/\partial i\psi_n^*$, with the Hamiltonian
\beA\label{eq:hamiltonian}
\mathcal{H} &=& \sum_n\mathcal{H}_n\nonumber\\
                   &\equiv& {\sum_n}[{\omega_n (p_n) +J(\psi_n^*\psi_{n+1}+\psi_n\psi_{n+1}^*)}].
\eeA

In the absence of damping, the system has the two conserved quantities $P=\sum_n p_n$ and $\mathcal{H}$, which give the two continuity equations for the local powers and energy:
\begin{subequations}
\begin{align}
\dot{p}_n= & j_{n+1}^M-j_{n}^M,\label{eq:dpdt}\\ 
\dot{\mathcal{H}}_n=& j_{n+1}^E-j_{n}^E\label{eq:dedt}. 
\end{align}
\end{subequations}

Those equations lead to the definition of the local magnetisation and energy currents, respectively
\begin{subequations}
\begin{align}
j_n^M = & 2J{\rm{Im}}\average{\psi_n\psi_{n+1}^*},\label{eq:mag}\\ 
j_n^E = & 2J{\rm{Re}}\average{\dot{\psi_n}\psi_{n+1}^*},\label{eq:energy}
\end{align}
\end{subequations}
where the chevrons indicate ensemble average.

When the system is dissipative, by conserved quantities we mean that they obey conservation equations that relate the time derivative of those quantities to the (energy and magnetisation) currents, the sources (thermal fluctuation and chemical potential) and the losses (damping). The currents are coupled in the sense of linear irreversible thermodynamics: a temperature difference generates a flow of energy and of spin. In a similar way, a difference in chemical potential generates both currents. The relation between forces and fluxes is described in the linear regime by the Onsager matrix, which can be used to calculate the energy and spin conductivity and the Seebeck coefficient \cite{iubini12,iubini13}. Although the currents are both due to the same (dipolar) coupling between the disks, they are in principle independent quantities. In particular, we remark that, while the particle current is proportional tho the correlation functions between the $\psi_n$, the energy current contains terms proportional to their time derivatives, and consequently to the frequency. This feature has been already observed in the seminal works on the off-equilibrium DNLS \cite{iubini12,iubini13,borlenghi14b}
In the phase-amplitude representations, the steady-states currents Eqs. (\ref{eq:mag}) and (\ref{eq:energy}) read 

\begin{subequations}
\begin{align}
j_n^M = & J\average{\sqrt{p_np_{n+1}}\sin[\Delta_{n,n+1}(t)+\beta]}\label{eq:mag2}\\
j_n^E = & J\omega_n(p_n)\average{\sqrt{p_np_{n+1}}\sin[\Delta_{n,n+1}(t)+\beta]}\label{eq:energy2} ,
\end{align}
\end{subequations}
with $\Delta_{n,n+1}(t)\equiv\phi_n(t)-\phi_{n+1}(t)$. The extra phase $\beta=\arctan\alpha$ stems from the condition of dissipative coupling previously discussed
\cite{iubini13,borlenghi15b,borlenghi16a}. 

The crucial observation here is that transport is a \emph{coherent} phenomenon, that occurs whenever the spin oscillators are phase-synchronised. In fact, the (unwrapped) phases grow in time as $\phi_n(t)\approx\omega_n\times t$, with
$\omega_n$ the frequency of the $n_{\rm{th}}$ oscillator. When the oscillators have the same frequencies, $\phi_n\approx\phi_{n+1}$, so that the currents
are proportional to $\sin\beta$. In the absence of phase synchronisation, the currents oscillate around zero and vanish in average~\cite{borlenghi14a,borlenghi14b}.

Note that the phase $\beta$ is given by the fluctuation-dissipation theorem Eq.(\ref{eq:fdt}), and determines the coupling strength with the bath. In the absence of dissipation, the system is conservative and
decoupled from the bath, thus no current can propagate through it. The condition $\beta=0$ guarantees that the currents vanish when the phase different is zero in the conservative case.

\section{micromagnetic simulations}%
The dependance of transport on the synchronisation  suggests the currents can be enhanced by applying a rf field, that increases both the local SW powers and the phase-locking in the system.

To investigate this issue, we have studied the dynamics of the chain by means of micromagnetics simulations, using the 
Nmag finite elements software~\cite{fischbacher07}. 
Each disk is represented by a tetrahedral mesh with maximum size of 3 nm, of the order of the Py exchange length, and was generated using
the Netgen package~\cite{schoeberl97}. The disks have radius $R=20$ nm, thicknesses $t=3$ nm, and they are separated by a distance $d=3$nm. 
The exchange stiffness $A=1\times 10^{-11}$ J/m is that of Py. The other micromagnetics parameters are the gyromagnetic ratio $\gamma=1.873\times 10^{11}$ rad$\times$s$^{-1}$ $\times$T$^{-1}$,
the saturation magnetisation $M_{s}=0.94$ T/$\mu_0$,  and the Gilbert damping parameter $\alpha=8\times 10^{-3}$ . Those parameters have been used in previous studies of nano-disks systems \cite{borlenghi15b} and are close to the
experimental parameters of Ref. \cite{naletov11}.

The rf pump field that appears in the DNLS equation Eq.(\ref{eq:dnls}) is implemented in the Nmag solver by adding the perturbation
$\bm{f}=(h_{\rm{rf}}\cos\omega_{\rm{rf}}\hat{\bm{x}},h_{\rm{rf}}\sin\omega_{\rm{rf}}\hat{\bm{y}},0)$ 
to the magnetisation vectors at each time step and mesh element, and is conveniently expressed in $M_s$ units.

The time evolution was computed for 110 ns with an integration time step of 1 ps. The relevant observables were time-averaged in the stationary state after a transient time of 50 ns, 
and then ensemble-averaged over 30 different realisations of the thermal field.
The finite temperature simulations were performed keeping the temperatures at the boundaries fixed as $T_+=15$ K and $T_-=5$ K, with $\Delta T=T_+-T_-=10$K.
Unless stated otherwise, we consider an applied static field $H_{\rm{ext}}=1$T.

The output of the simulations consists of the magnetisation vectors $\graffb{\bm{M}^n({\bm{r}}_n,t)}$, $n=1,...,10$. Each vector depends on the mesh node coordinate ${\bm{r}}_n$ inside the $n_{\rm{th}}$ disk. 
The collective magnetisation dynamics of each disk is given by the volume average \cite{slavin09,borlenghi15b}, 
\be\label{eq:mav}
\bm{M}^n(t)=\frac{1}{V}\int_{V_n} \bm{M}^n({\bm{r}}_n,t){\rm{d}}^3\bm{r}_n
\ee

from which the SW amplitudes $\psi_n(t)$s and the currents Eqs.(\ref{eq:mag}) and (\ref{eq:energy}) are computed. Note that the value of the coupling $J$ cannot be extracted from our simulations, so that the currents
are expressed per unit coupling and are pure numbers.

\begin{figure}[t]
\begin{center}
\includegraphics[width=9cm]{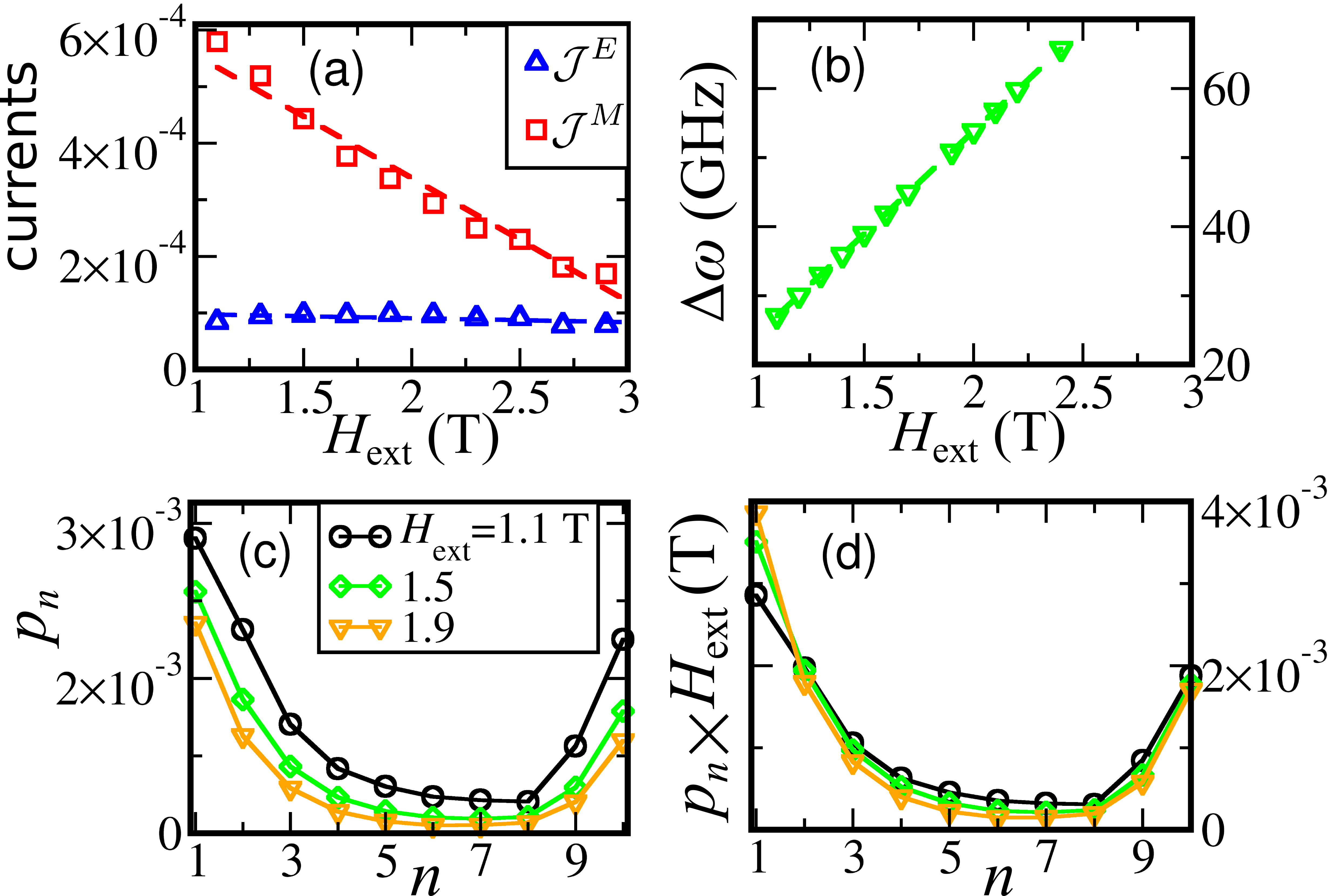}
\end{center}
\caption{a) Net currents vs the static field $H_{\rm{ext}}$. At increasing field, $\mathcal{J}^M$ decreases while $\mathcal{J}^E$ remains unchanged.
b) Frequency shift of the spectrum $\Delta\omega$ vs the static applied $H_{\rm{ext}}$. The dashed lines are linear fits. c) SW power profiles computed for different values of the applied field.
d) SW power profiles rescaled by the field, which overlap in the bulk. The lines are guides to the eye.} 
\label{fig:figure3}
\end{figure}

Here, we discuss the effect of the static field as a means to control separately the currents.
In a chain without dissipation in the bulk, transport is described by the total currents $j^{M/E}=\sum_{n}j_n^{M/E}$ that flow through the system~\cite{lepri03,iubini12}.
In the present case, part of the local currents is dissipated in the bulk, and one needs a different quantity to describe transport.
The relevant quantities here are the net currents that flow through the system, $\mathcal{J}^{M/E}=j^{M/E}_2+j^{M/E}_8$.
Those correspond to the two currents injected from baths, minus the SW power/energy dissipated in the first and last disk \cite{borlenghi15b}.

 as it can be seen from Eqs.(\ref{eq:mag2}), and (\ref{eq:energy2}).
Fig. \ref{fig:figure3}a) shows the net currents as a function of $H_{\rm{ext}}$, with zero pump field and $\Delta T=10$K. Increasing the static field, one can see that  $\mathcal{J}^M$ decreases linearly while $\mathcal{J}^E$ remains constant.
To describe this effect, it is useful to write the ensemble-averaged equation for the $p_n$s in the presence of dissipation \cite{borlenghi14b}.
\be\label{eq:powercons}
\frac{d}{dt}\average{p_n}=-2\Gamma_n\average{p_n}+2D_nT_n+j^M_{n+1}-j^M_n.
\ee
Here damping and thermal fluctuations act respectively as sink and sources for the local SW powers. In the steady state, one has
$d\average{p_n}/dt=0$, so that Eq.(\ref{eq:powercons}) becomes 
\be\label{eq:steadyp}
\average{p_n}=\frac{1}{\omega_n}\left(T_n+j^M_{n+1}-j^M_n\right). 
\ee
Since the local frequencies are proportional to the applied field (see Fig.\ref{fig:figure3}b)), increasing the latter
suppresses both the source term $T_n/\omega_n$ and the currents, reducing the magnitude of the $p_n$. 
The magnetisation current Eq.(\ref{eq:mag}) is proportional to $\sqrt{p_np_{n+1}}$, and should decrease with the applied field.
On the other hand, the energy current Eq.\ref{eq:energy} is proportional to the product $\omega_n\times\sqrt{p_np_{n+1}}$ and remains constant at increasing applied field.  

This qualitative picture is corroborated by confronting the SW power profiles in Fig.\ref{fig:figure3}c) and d). If the powers decrease as $1/|H_{\rm{ext}}|$, one expects that the quantities $p_n \times H_{\rm{ext}}$ remains constant at increasing field.
This feature is displayed in Panel d), which shows that current profiles multiplied by the static field overlap in the bulk.

Next, we address the effect of the rf field on the transport. The power spectrum of the system, which reveals the SW modes, is given by the absolute value of the Fourier transform of the total SW amplitude $\Psi=\sum_n\psi_n$.
The zero temperature spectrum, thoroughly described in Ref.[\onlinecite{borlenghi15b}], is reported in Fig.\ref{fig:figure1}b). It consists of five dipolar modes $(\omega_1,...,\omega_5)$ with frequencies respectively 
$(18,10.7,20.5,21.8,24)$ GHz. Those modes correspond to excitations localised in different parts of the chain. In particular, the mode $\omega_1$ corresponds to the precession of the first and tenth disk, 
the mode $\omega_2$ to the second and ninth disk, and so on until the mode $\omega_5$ which is associated to the precession of the two central disks. Note that the different height of the peaks depends on initial conditions \cite{naletov11} and is not related to transport.
We will see in particular that the smallest mode $\omega_1$ is the one that most contribute to transport.
\begin{figure}[t]
\begin{center}
\includegraphics[width=9cm]{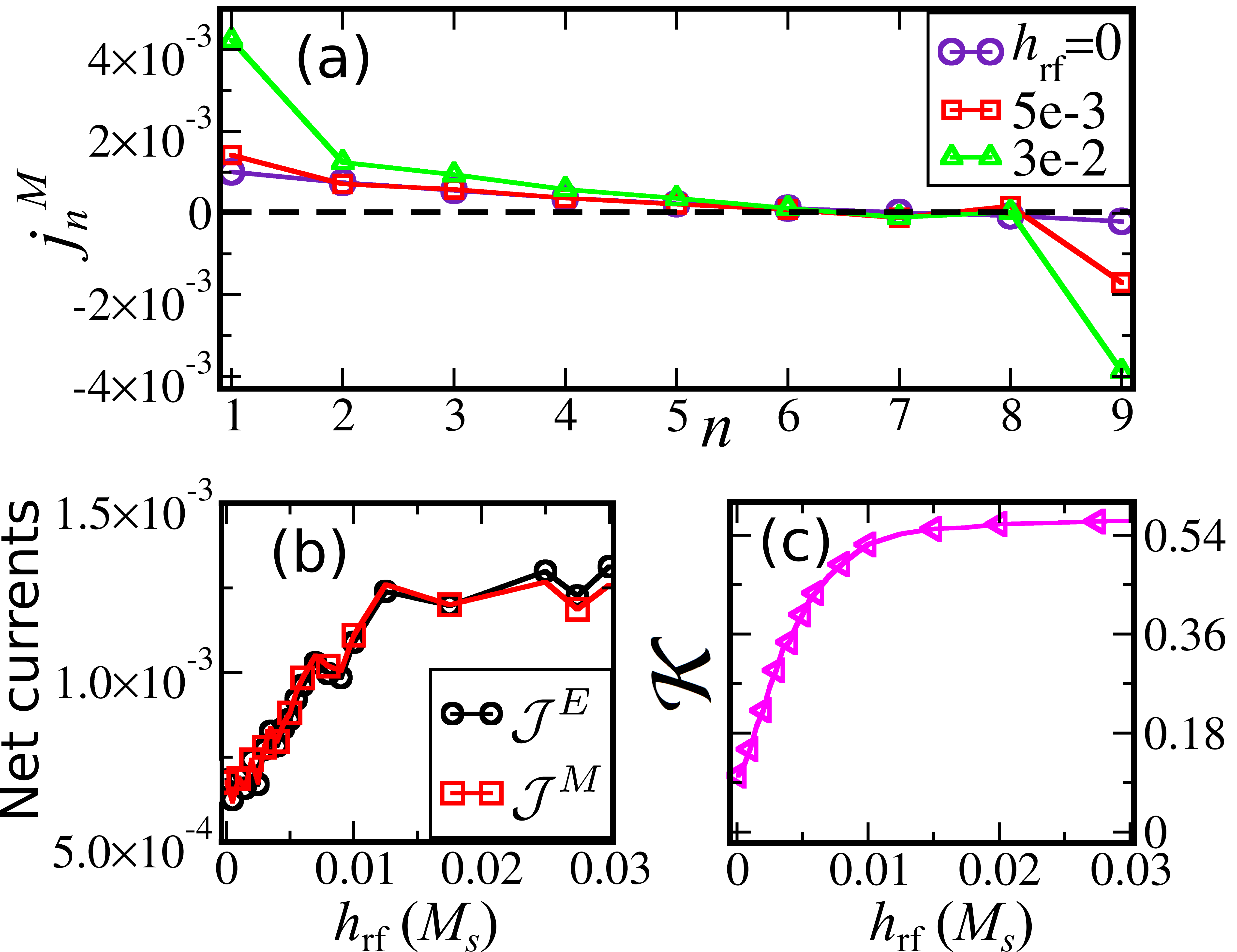}
\end{center}
\caption{a) Profiles of the magnetisation currents $j_n^{M}$, computed for different values of the rf field. b) Net currents and c) 
Kuramoto parameter vs the rf field. The lines are guides to the eyes.} 
\label{fig:figure2}
\end{figure}

The crucial feature here is that each SW mode can be excited \emph{selectively} by applying a rf field with the corresponding frequency, allowing to localise energy in different parts of the chain. This can be seen in Fig.\ref{fig:figure1}c), that shows the
SW power profiles at zero temperature, in the presence of a rf field with amplitude $h_{\rm{rf}}=5\times10^{-3} M_s$ and frequencies $\omega_{\rm{rf}}=(\omega_1,\omega_3,\omega_5)$. 
Although the strength of the rf field is uniform along the chain, the precession amplitude increases the most in the disks with frequencies that resonate with the field.
In particular, when $\omega_{\rm{rf}}=\omega_{2,3}$ the precession occurs mostly in the center of the chain, while when $\omega_{\rm{rf}}=\omega_1$, the edge modes are excited and the precession occurs mostly in the first and last disk.

We mention also that we have tested the effect of the rf field with frequencies of the other SW modes. It has been found that the conductance does not improve in those cases. In particular, we have observed that the dynamics becomes chaotic
and the phase coherence is disrupted.

Exciting the edge modes has the effect of increasing the powers $p_n$, and thus the spin-chemical potential~\cite{borlenghi14b}, in the first and last disk of the chain. 
This together with the fact that the system becomes more coherent, should increase the thermal and spin conductivity.
To verify this, we have computed the time evolution in the presence of both thermal gradient and the rf pump field 
with frequency $\omega_{\rm{rf}}=\omega_1=18$GHz and intensity $h_{\rm{rf}}$ ranging between zero and $3\times 10^{-2}M_s$.

The profile of the magnetisation current $j_n^M$ is shown in Fig.\ref{fig:figure2}a). Positive (resp. negative) currents propagate towards the left  (resp. right). Energy currents have the same profiles as $j_n^M$ up to a scaling factors and therefore
they are not reported. At increasing field, one can observe a strong increase of the currents at the edges and of the positive currents in the bulk.

The net currents, plotted in Fig.\ref{fig:figure2}b) versus $h_{\rm{rf}}$, increase linearly until $h_{\rm{rf}}=0.01$, and then reach a plateau.
Both currents have the same profile up to a scaling factor ($\mathcal{J}^E$ is magnified by a factor 8 for better visibility).

Synchronisation is described by the  Kuramoto order parameter \cite{kuramoto75}
\be\label{eq:kuramoto}
\mathcal{K}=\frac{1}{N}\left|\average{\sum_n e^{i\phi_n}}\right|.
\ee
This quantity ranges from 0 for a completely incoherent state to 1 for a completely phase-synchronised one. In Fig. \ref{fig:figure2}c), one can see that $\mathcal{K}$ increases with the rf field up 
to a maximum value of 0.54. The quantities $\mathcal{J}^{M/E}$ and $\mathcal{K}$ have similar profiles, and they both increase linearly up to $h_{\rm{rf}}=0.01M_s$, indicating the connection between transport and synchronisation.
Note that, although the coherence increases significantly, the system does not become completely synchronised even at high field. This is due to two well known phenomena. At first, in this kind of systems the oscillators are repulsive, in the sense that the lower energy modes correspond to an anti-phase precession between the magnetisation of the disks \cite{slavin09,naletov11,borlenghi14a}. This kind of system cannot synchronise completely, no matter the strength of the driving field. Then, the frequency band of mutual phase-locking is non-linear. The nonlinearity introduces a phase shift between the driving signal and the oscillators that tends to suppress mutual phase-locking \cite{slavin09}.

\section{conclusions} %
In summary, we have described nanoscale control of heat and spin conductance in an artificial spin chain using the applied static field and radio frequency pump field. Our simulations show that the rf field acts by increasing the chemical potentials at the edges
of the system and promoting the phase coherence between the spin-oscillators. Furthermore, by increasing the static field, we observe that the magnetisation current is suppressed, while the energy current remains constant. 
These features were described by a very general oscillator model, the non-equilibrium DNLS equation.
The generality of the DNLS model elucidates the connection between synchronisation and transport, and suggests that the same mechanism could be used in a variety of physical systems.

\acknowledgements %
We thank Prof. Andrei Slavin, Dr. Stefano Iubini and Dr. Stefano Lepri for useful discussions. This research was supported by the Stiftelsen Olle Engkvist Byggm\"astare. 
We also acknowledge financial support from Vetenskapsradet (VR), The Royal Swedish Academy of Sciences (KVA), the Knut and Alice Wallenberg Foundation (KAW), Swedish Energy Agency (STEM), Swedish Foundation for Strategic Research (SSF), Carl Tryggers Stiftelse (CTS), eSSENCE, and G\"{o}ran Gustafssons Stiftelse (GGS).
The computations were performed on resources provided by the Swedish National Infrastructure for Computing (SNIC) at the National Supercomputer Center (NSC), Link\"oping University and at the PDC center for high-performance computing, KTH.





\end{document}